\documentclass[12pt]{iopart}
\usepackage{graphicx}
\usepackage{multirow,tabularx}
\usepackage{subcaption}

\begin{document}

\title[]{Robust Errant Beam Prognostics with Conditional Modeling for Particle Accelerators}

\author{Kishansingh Rajput$^{*, 1, 3}$, Malachi Schram$^1$, Willem Blokland$^2$, Yasir Alanazi$^1$, Pradeep Ramuhalli$^2$, Alexander Zhukov$^2$, Charles Peters$^2$, Ricardo Vilalta$^3$}
\address{$^1$ Thomas Jefferson National Accelerator Facility, Newport News, VA 23606, USA}
\address{$^2$ Oak Ridge National Laboratory, Oak Ridge, TN 37830, USA}
\address{$^3$ Department of Computer Science, University of Houston, Houston, TX 77204, USA}
\ead{$^*$ kishan@jlab.org}

\vspace{10pt}

\begin{indented}
\item[]November 2023
\end{indented}

\begin{abstract}
Particle accelerators are complex and comprise thousands of components, with many pieces of equipment running at their peak power.
Consequently, particle accelerators can fault and abort operations for numerous reasons.  
These faults impact the availability of particle accelerators during scheduled run-time and hamper the efficiency and the overall science output. 
To avoid these faults, we apply anomaly detection techniques to predict any unusual behavior and perform preemptive actions to improve the total availability of particle accelerators.  
Semi-supervised Machine Learning (ML) based anomaly detection approaches such as autoencoders and variational autoencoders are often used for such tasks. However, supervised ML techniques such as Siamese Neural Network (SNN) models can outperform unsupervised or semi-supervised approaches for anomaly detection by leveraging the label information.
One of the challenges specific to anomaly detection for particle accelerators is the data's variability due to system configuration changes.
To address this challenge, we employ Conditional Siamese Neural Network (CSNN) models and Conditional Variational Auto Encoder (CVAE) models to predict errant beam pulses at the Spallation Neutron Source (SNS)  under different system configuration conditions and compare their performance. We demonstrate that CSNN outperforms CVAE in our application. 
\end{abstract}

%
\noindent{\it Keywords}: Anomaly detection, Siamese Model, Conditional Siamese Model, Variational Autoencoder, Conditional Variational Autoencoder, CSNN, CVAE, Spallation Neutron Source, Particle Accelerators, Anomaly Prediction, Prognostics
%
%
%
%
\section{Introduction}  
The Spallation Neutron Source (SNS) accelerator facility at Oak Ridge National Laboratory (ORNL) delivers a 60 Hz pulsed 1.1 GeV proton beam at 1.7 MW, making it the world's highest power proton accelerator. 
The beam is accelerated in a super-conducting linear accelerator and accumulated in a ring to form a very short and intense pulse with an intensity of up to $1.6 \times 10^{14}$ protons. 
The protons are then directed to a stainless steel vessel filled with liquid mercury where the impact spalls the mercury atoms and neutrons are released \cite{2014Henderson-SNSSystemDesign} for material research. 
High availability of the SNS accelerator (greater than 95\%)  is challenging to attain, even with judicious maintenance and monitoring, since errant beam pulses still occur and cause downtime~\cite{Peters:2018uvp}. 
To reduce downtime, radio-activation of beam line elements, and damage to components, we explored ML methods that use data from existing diagnostic sensors to predict equipment failures and errant beams. 
Previous studies at the SNS~\cite{2020-Rescic-AccelFailure, UQ_SNS}, show that precursor information is present in the available data, which can predict an impending fault. 

Existing accelerator data provides a large amount of normal samples and a small number of anomaly samples.
As such, researchers have been resorting to semi-supervised learning methods such as autoencoder (AE)~\cite{10.5555/104279.104293} or Variational AE (VAE)~\cite{https://doi.org/10.48550/arxiv.1312.6114} to leverage a large amount of existing normal data samples~\cite{AE_slac, VAE_Slac, pol2020anomaly}.
AE and VAE models learn to reproduce the normal data samples and predict anomalies based on the reconstruction error (the error between the input data and the corresponding reproduced data).
However, our previous study~\cite{UQ_SNS} on this application concluded that Siamese neural network (SNN), a supervised learning method, can outperform AE.
SNN can also leverage many normal samples and avoid label imbalance as described in Section~\ref{ch:method}.

Particle accelerators have multiple configuration parameters updated throughout operation for control and optimization.
The changes in the configurations impact the equipment behavior and, subsequently, the measurements from the diagnostic sensors, making anomaly detection more challenging.
Machine Learning (ML) models trained on the data from a particular beam configuration may no longer be valid for the following accelerator configuration.
Conditional models can tackle such shifts in the data distribution by using the beam configuration parameters as conditional inputs.
In this research, we study the performance of conditional-SNN (CSNN) and conditional-VAE (CVAE)\cite{CVAE} for predicting impending errant beam pulses using the beam current data from eight different SNS accelerator beam configurations. 
The results indicate that the CSNN model trained on data from all eight beam configurations outperforms dedicated SNN models trained on data from a single beam configuration.
Based on the specific cases studied in this paper, we show that the CSNN outperforms the CVAE which is also trained on data from all beam configurations.

In Section~\ref{ch:previous}, we summarize previous research in anomaly detection, classification, and prognostication as it applies to particle accelerator use cases.
In Section~\ref{ch:data_description}, we describe the data source, collection, and curation procedures.
Section~\ref{ch:method} describes the ML methods we explored for this paper.
In Section~\ref{ch:results}, we present the comparison results between the SNN, CSNN, and CVAE .
In addition, we present the model architecture and hyper-parameter optimization process.
Finally, in Section~\ref{ch:conclusion}, we present our conclusions and outlook.

\section{Previous Work} 
\label{ch:previous}


Anomaly detection problems have been studied with statistics and visualization-based methods for many years.
Methods such as K-nearest neighbor clustering~\cite{KNN, KNN2}, quantile method or whisker plots~\cite{WhiskerPlots}, probabilistic envelope-based approach~\cite{rajput2022probabilistic} have been applied for anomaly detection in different fields.
Recent advancements in ML have been a good fit for many applications, including anomaly prediction.
Deep Learning (DL), particularly,  is very expressive and can handle large and complex data sets;
due to high expressivity and a large number of parameters, it can model large and complex systems.


The use of ML for particle accelerator applications has grown in recent years including, but not limited to, diagnostics~\cite{Emma:2018meg, Sanchez-Gonzalez:2016zhm,Wielgosz:2016xhl,Scheinker:2015mra, alex2020advanced, PhysRevLett.121.044801,info12030121}, anomaly detection/forecasting/classification~\cite{UQ_SNS,Rescic:2020ueu,RESCIC2022166064,tennant2020superconducting,Powers:2019ioo}, and optimization/controls~\cite{miskovich2023multipointbax, 9667198, pmlr-v162-kaiser22a, PhysRevAccelBeams.25.062802, PhysRevAccelBeams.24.104601,Kafkes:2021jse, Edelen:2016dqu, hirlaender2020modelfree}.
\\

AEs and VAEs are commonly used for anomaly detection, where the samples used during training are assumed to be normal.
Previous work using AEs for particle accelerator sub-systems include the Air Conditioning systems~\cite{osti_1996207} and magnet faults in the Advanced Photon Source storage ring at Argonne National Laboratory~\cite{AE_slac}.
Recently, VAEs have been applied for High Voltage Converter Modulator (HVCM) anomaly detection at SNS accelerator~\cite{ALANAZI2023100484,RADAIDEH2022103704}.
There have also been methods proposed, such as ~\cite{VAE_Slac}, to deal with impurities in the normal training data for VAEs.
\\

In addition, there have been advances in research to understand uncertainty quantification (UQ) from ML models.
New accelerator physics studies have included UQ to estimate out-of-distribution uncertainties using data-driven ML-based surrogate models.
Examples include developing a UQ-based surrogate model for a cyclotron-based model~\cite{doi:10.1137/16M1061928},  modeling the FNAL Booster accelerator complex for a reinforcement learning application~\cite{Kafkes:2021jse}, and uncertainty aware anomaly predictions~\cite{UQ_SNS}.
Furthermore, recently published studies compared Bayesian Neural Networks (BNN), ensemble methods, and a novel deep Gaussian process approximation method for particle accelerator applications~\cite{PhysRevAccelBeams.24.114601, PhysRevAccelBeams.26.044602,  rajput2023uncertainty, goldenberg2023distance}.
These new techniques will allow us to leverage the strength of Deep Neural Networks (DNNs) while providing some safety guarantees.
\\

Although these studies show very promising results, they typically do not address how to apply them for longer periods of time with varying operational conditions.
Presented in this paper is a study on conditional ML to address the underlying changes in the particle accelerator configurations.

\section{Data Description}
\label{ch:data_description}
There are various sensors deployed at SNS including Beam Positional Monitors (BPM), Beam Loss Monitor (BLM), and Differential Current Monitor (DCM). 
These sensors provide data to perform diagnostics and preemptive maintenance.
This research uses beam current data from DCM as described below.

\subsection{Data Source and Collection}

\begin{figure}[h]
    \centering
    \includegraphics[width=0.7\textwidth]{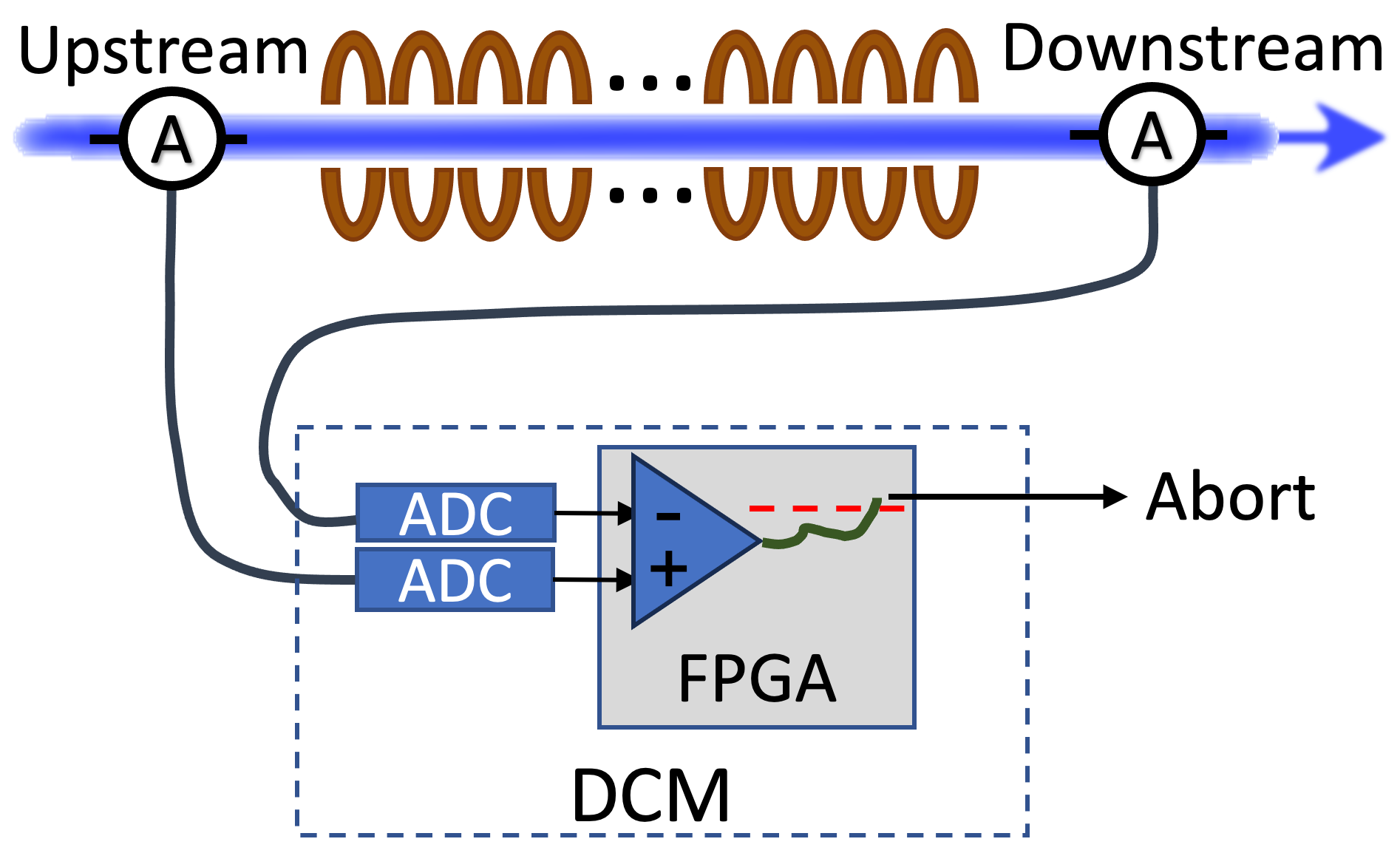}    
    \caption{Differential Current Monitor (DCM) along with beam demonstration}
    \label{fig:DCM}
\end{figure}

SNS accelerator employs a DCM~\cite{dcm-1.1, dcm-2.2} to protect the Super Conducting Linac (SCL) from adverse effects of beam losses~\cite{KIM201720}. 
Figure~\ref{fig:DCM} shows a schematic diagram of the DCM.  
It continuously monitors upstream and downstream beam current waveforms to detect any beam loss across the SCL. 
When significant beam loss is detected, it aborts the beam to minimize possible damage or activation. 
DCM uses a Field Programmable Gate Array (FPGA) and a dedicated communication line with the Machine Protection System (MPS) to enable fast aborts. 
Initially, the DCM only streamed a subset of the waveforms to a server for storage.
At regular intervals, normal waveforms were stored and waveforms right before, during, and right after an errant beam pulse. 
All waveforms of all beam pulses are streamed to an edge computer for inferences of models and to store the data. 

With this study, we aim to predict errant beam pulses before they occur using normal prefault pulses. 
Our previous studies ~\cite{UQ_SNS, 2020-Rescic-AccelFailure} have shown that there are precursors in the normal pulse to indicate an impending errant beam pulse.
To achieve this, the DCM was programmed to archive beam current waveforms preceding errant beam pulses and regularly archive waveforms during normal operation. 
The prefault normal waveforms are labeled as anomalies and waveforms from the normal operation are labeled as normal (reference) for the ML study.

We also recorded a snapshot of beam configuration settings along with the beam current data for conditional modeling.
Eight relevant beam tuning parameters are considered for this study, namely, Ramp-Up-Pulse-Width, Ramp-Up-Pulse-Width-Change, Ramp-Up-Width, Beam-On-Width, Beam-On-Pulse-Width, Ramp-Down-Pulse-Width, Gate-Source-Offset, and Beam-Ref-Gate-Width. The description of these parameters is included in the Appendix.

Figure \ref{fig:sns-waveforms4} shows an example of the beam current waveform. 
The top plot shows a series of macro-pulses, a 1 ms long pulse repeated at 60 Hz. 
This macro-pulse consists of approximately 1000 mini-pulses. 
Each mini-pulse is $\sim$650 ns, followed by a gap of $\sim$350 ns. 
Within each mini-pulse are the micro-pulses, not shown, which are the Radio Frequency (RF) buckets filled with the beam particles spaced at 402.5~MHz.
The bottom plot shows a beam pulse's current waveform with the initial ramp-up in intensity in the beginning of the macro-pulse, as well as the different widths of the mini-pulses during the macro-pulse. 
This setup is typical for a production-style beam. 

\begin{figure}[h]
    \centering
    \includegraphics[width=0.7\textwidth]{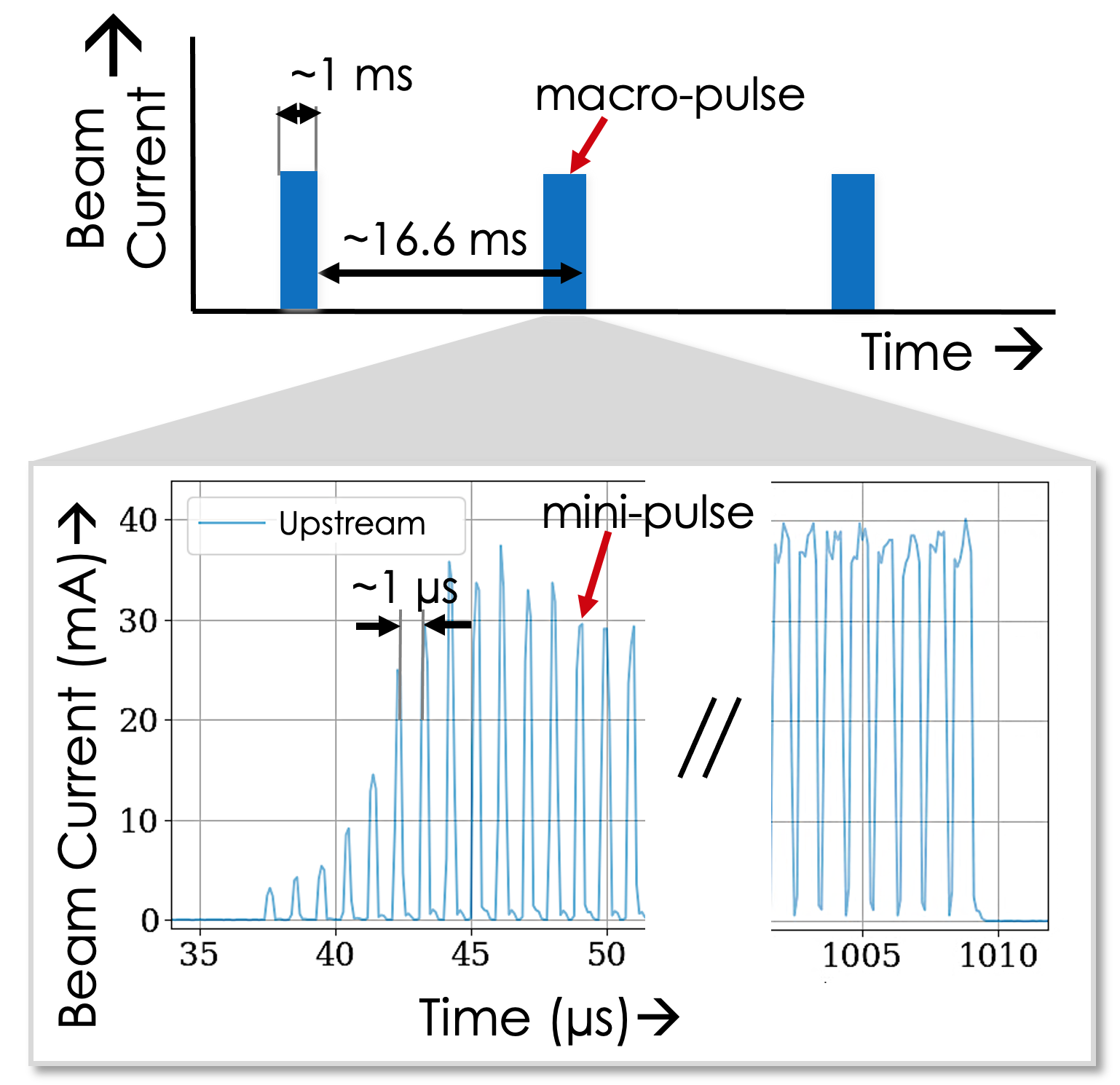}    \caption{Beam macro-pulse pattern and beam current waveform showing the mini-pulses}
    \label{fig:sns-waveforms4}
\end{figure}

The waveform has a length of about 1.2 ms and is digitized at 100 MS/s. The beam pulse's timestamp, beam configuration parameters, and cycle ID are collected and archived along with the upstream and downstream beam current waveforms. 

\subsection{Training Data Generator}

Each macro pulse comprises $\sim$120K time steps; however, for this study, we only use 10K time steps (from time step 3K to 13K) as our research indicated that the pre-cursors for an upcoming errant beam pulse are prevalent in this region.
Furthermore, a peak counter approach was employed to exclusively select beam current waveforms associated with neutron production and avoid outliers such as test beams or beam studies. 
This technique leverages the ``find peaks'' method in the SciPy library~\cite{SciPy}. 
The parameters for peak identification were set with a minimum height threshold of 2 mA and a stipulated minimum distance of 75 steps, equivalent to 750 ns, between adjacent peaks.
The waveforms of the macro-pulse must encompass a minimum of 900 mini-pulses and exhibit a consistent repetition frequency of 60 Hz to indicate that these pulses were produced during full power beam and not during physics studies or testing. 
These strict requirements effectively filter out any potential outliers in accelerator conditions.

The dataset is divided into three parts, specifically training (64\%), validation (16\%), and testing (20\%) according to their beam setting configurations. 
All data collected during normal operation was considered normal data.
For SNN training, since the model requires two input samples, we developed a training data generator using tensorflow data generator~\cite{tf} to create combinations of normal-normal data labeled as 0 and normal-prefault data labeled as 1. 
The data generator has two hyper-parameters: the number of anomalies per batch and the samples per anomaly. 
The former controls the number of unique prefault beam current samples in a batch, and the latter controls the number of normal beam current waveforms paired with each anomaly.
The number of normal-normal samples matches the number of normal-prefault samples per training batch in the data generator.
It is important to note that only samples from the same beam configuration are paired with each other.
The data generator maintains separate counters for both normal and anomaly waveforms for each configuration setting to iterate over them in sequential order during training and evaluation.
To introduce randomness, the normal and anomaly data are shuffled at the end of each epoch.

\section{Methods}
\label{ch:method}
As discussed in Section~\ref{ch:previous}, there are various techniques to predict anomalies. 
In this study, we compare VAEs and SNNs with their conditional variants to accommodate data from different beam configurations. 
Below, we provide a brief background of these techniques.

\subsection{Variational Autoencoders}
VAEs~\cite{https://doi.org/10.48550/arxiv.1312.6114} are commonly used for anomaly detection~\cite{vae_anomaly, ALANAZI2023100484} when there are few anomaly examples. 
VAEs are normally used for semi-supervised learning and can address imbalanced classification problems. 
VAEs are trained only on normal data to model the normal behavior of the system. 
The model identifies any anomalous data using the reconstruction error at inference time. 
A typical VAE consists of two Artificial Neural Networks (ANNs), an encoder $\Psi$, and a decoder $\Phi$. 
The $\Psi$ maps input space $x$ into a reduced representation $z$ that is forced to follow a specific well-known distribution, such as a Gaussian distribution. 
The latent $z$ is generated by sampling from $\mu$ and $\sigma$ parameters estimated by the $\Psi$. 
A new parameter $\epsilon$ is also introduced to reparametrize the sampling layer $z$ and allow the model to backpropagate through the entire network.
The latent variable is then fed to the $\Phi$ to reconstruct the input data.  
The loss function for a VAE is motivated by variational inference~\cite{variationalinference} by minimizing the Kullback–Leibler divergence (KLD)~\cite{kullback1951information} between the posterior $p(z|x)$ and the encoded prior distribution $q(z) = \mathcal{N}(0,1)$:
\begin{equation} \label{eq:2}
    \mathcal{L}_{\rm VAE}= \|x-\Phi(\Psi(x))\|^2 +\eta~{KLD}~(q(z) ~||~ p(z|x))
\end{equation}
where the first term is the reconstruction error, the second term computes KLD, and $\eta$ is the harmonic parameter to balance the two.

\subsection{Siamese Neural Network Model}
\label{sec:siamese-model}

\begin{figure}[h]
    \centering
    \includegraphics[width=0.5\textwidth]{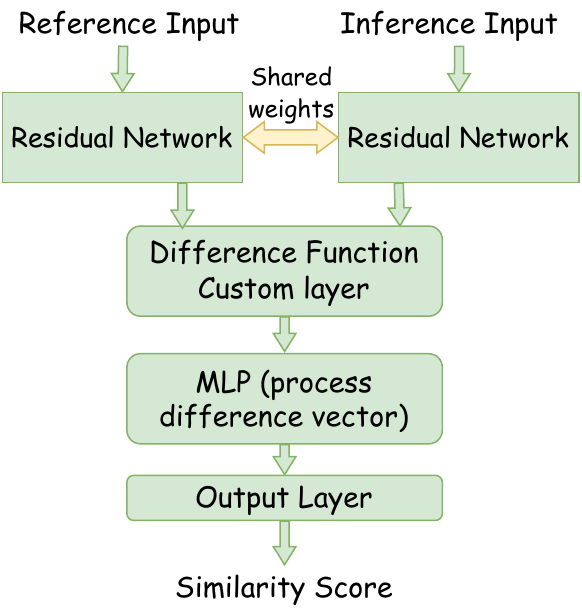}    
    \caption{The SNN model architecture comprises two input layers for the reference waveforms and the acquired waveform, followed by one common ResNet model and a distance function. Then a MLP block with dropout layers is applied to avoid over-fitting, and finally, an output layer is used to compress the dimensionality and produce the similarity scalar.}
    \label{fig:SiamesePlot}
\end{figure}

The SNN model takes a different approach to anomaly detection than traditional classification methods, or VAEs, by learning the similarities between a pair of inputs. 
Instead of focusing on classifying individual data samples, it seeks to capture similarities (or dissimilarities for anomalies).
This is achieved by encoding two separate input vectors using twin networks (sharing weights and biases) and then comparing the reduced latent space using a distance metric, as illustrated in Figure~\ref{fig:SiamesePlot}. 
The twin networks are deployed to learn and extract the relevant features for similarity prediction between normal-normal and normal-abnormal combinations.
As it learns the relevant features, it produces a reduced representation of the original inputs that are then compared using a 
Lambda difference layer as described in Equation~\ref{eq:lambda_difference}.
\begin{equation}
D(h_{x_1}, h_{x_2}) =  \mid h_{x_1}^{2} - h_{x_2}^{2} \mid
\label{eq:lambda_difference}
\end{equation}
Here, $h_{x_1}$ and $h_{x_2}$ are reduced representations of reference and inference inputs, respectively.
The output from the difference layer is fed into a multi-layer perceptron (MLP) block with dropouts to avoid over-fitting before the final output layer that produces scalar similarity score between the two inputs.

During training, predicted similarity scores are compared with the ideal scores (labels) using a modified contrastive loss function described in Equation~\ref{eq:contrastive_loss}.

\begin{equation}
L(y,\hat{y}) =    \alpha \times (1-y)*\hat{y}^2 + y * (\max(\beta-\hat{y},0))^2
\label{eq:contrastive_loss}
\end{equation}

Here, $y$ is the true label, $\hat{y}$ is the predicted similarity score, $\alpha$ (set to 5.2 for this study) is a tuning parameter used to emphasize the similar pulses, and $\beta$ is another tuning parameter used to emphasize dissimilar pulses (set to 1 for this study). 

This study uses a ResNets~\cite{He2016DeepRL} model for the twin network with three stacked convolution residual blocks. 
Each residual block consists of skip connections between two convolution layers to improve the information propagation in the deep model.
Each ResNet block consists of a convolution layer, Rectified Linear Unit (ReLU) activation layer, a batch-normalization layer, and a dropout layer. 
The model architecture parameters are provided in Table~\ref{tab:model_architectures}.
\subsection{Conditional Models}
Conditional models learn a relationship between input and output data under given conditions. 
These conditions can be explicit features or variables influencing the relationship between inputs and outputs. 
In other words, the model's predictions depend on specific conditions or contextual information.
We employed conditional models that take beam configurations as input to an orthogonal MLP to handle the beam current data belonging to different beam configurations.
We compared both CSNN and CVAE for our application.

\begin{figure}[h]
    \centering
    \includegraphics[width=0.65\textwidth]{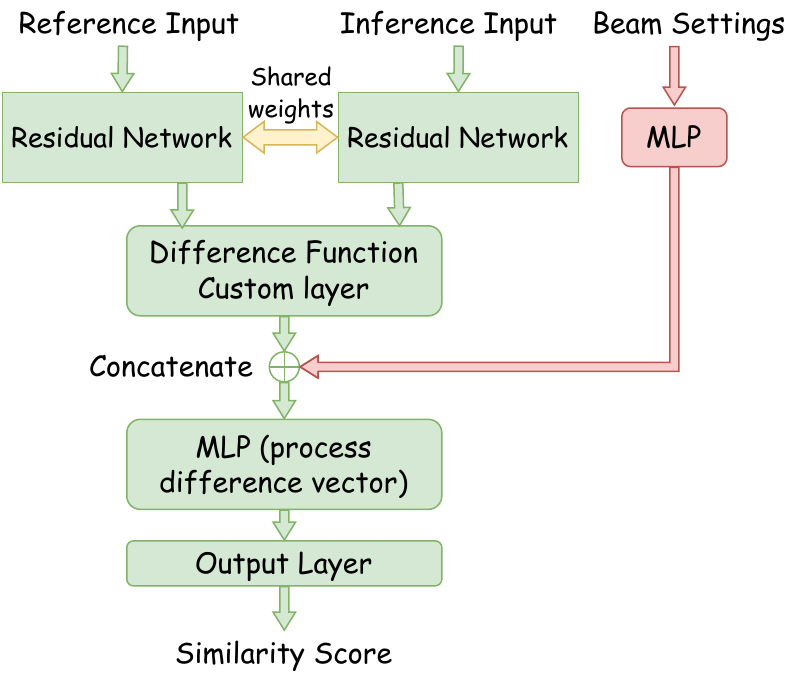}    \caption{The CSNN model architecture, similar to the SNN model with the addition of a third input for beam configurations and an MLP block for the embedding of the configuration vectors. The configuration embeddings are then concatenated with the latent difference before passing to the common MLP block followed by the output layer.}
    \label{fig:conditional-siamese-plot}
\end{figure}

CSNN consists of the same architecture described in Section~\ref{sec:siamese-model} with an additional conditional input.
The conditional input is then passed to an embedding sub-model as shown in red in Figure~\ref{fig:conditional-siamese-plot}. 
The configuration embedding MLP consists of 3 dense layers with 64, 32, and 16 nodes respectively along with leaky relu activations and 5\% dropouts.
The embedding transforms the beam configuration values into a latent space representation, which allows the model to learn the beam configuration landscape and leverage any correlations.
The embedded beam configuration is concatenated to the output from the difference layer as shown in Figure~\ref{fig:conditional-siamese-plot}.

CVAE~\cite{CVAE} was proposed to make diverse predictions for different input samples. A CVAE is a type of VAE that introduces a conditional variable into both the encoder and the decoder.
The objective function of the VAE can be modified by adding the variable $c$,
\begin{equation} \label{eq:3}
    \mathcal{L}_{\rm CVAE}=\|x-\Phi(\Psi(x))\|^2+\eta~{KLD}~(q(z,c)~ ||~ p(z|x,c))
\end{equation}
where we condition all of the distributions with $c$. CVAEs are an extension of VAEs by adding the conditional part in the $encoder$ and $decoder$ to associate the input samples with labels.
Therefore, at inference time, we have more control to generate samples that belong to specific conditions in contrast to a VAE that does not have control over the generated samples. 

All the models in this study are implemented using Tensorflow 2.8 \cite{tf} with Adam optimizer~\cite{Adam} for training. 
For efficient optimization, we used a learning rate decay (callback provided by tensorflow) of 85\% using the validation loss plateau with a patience value of 5 epochs.
To automatically tune the number of epochs required for convergence, we used an early stopping callback at the validation loss plateau with a patience of 20 epochs.



\section{Results}
\label{ch:results}
The beam current waveforms belonging to eight different beam configuration instances were used for the training and evaluation of the SNN, CSNN, and CVAE models, and the results are discussed in this section.

\begin{table}
\caption{\protect\label{tab:results}Comparison of the TPR at FPR of 0.1\% for the CVAE, SNN and CSNN models. The range shows the maximum and minimum from the ensemble models.}
\lineup
\centering
\begin{tabular*}{\textwidth}{@{}l*{15}{@{\extracolsep{0pt plus
12pt}}r}}
\br
\multirow{2}{*}{Config Id} & \multirow{2}{*}{CVAE} & \multirow{2}{*}{SNN} & \multirow{2}{*}{CSNN} & \multicolumn{2}{c}{\% improvement by CSNN} \\
 & & &  & over CVAE & over SNN \\
 
\mr
1 &  $0.233\pm^{0.100}_{0.167}$ & $0.133\pm^{0.333}_{0.133}$ & $\textbf{0.400}\pm^{0.067}_{0.000}$ & 71.67\% & 200.75\%\\
2 &  $0.316\pm^{0.000}_{0.053}$ & $0.368\pm^{0.053}_{0.263}$ & $\textbf{0.368}\pm^{0.105}_{0.000}$ & 16.45\% & 0.00\%\\
3 &  $0.217\pm^{0.028}_{0.075}$ & $0.394\pm^{0.008}_{0.008}$ & $\textbf{0.394}\pm^{0.008}_{0.244}$ & 85.51\% & 0.00\%\\
4 &  $0.000\pm^{0.000}_{0.000}$ & $0.146\pm^{0.188}_{0.104}$ & $\textbf{0.292}\pm^{0.083}_{0.125}$ & - & 100.00\%\\
5 &  $0.155\pm^{0.000}_{0.052}$ & $0.121\pm^{0.086}_{0.121}$ & $\textbf{0.207}\pm^{0.034}_{0.017}$ & 33.54\% & 71.07\%\\
6 &  $0.214\pm^{0.000}_{0.143}$ & $0.143\pm^{0.143}_{0.143}$ & $\textbf{0.357}\pm^{0.214}_{0.071}$ & 66.82\% & 149.65\%\\
7 &  $0.177\pm^{0.016}_{0.016}$ & $0.290\pm^{0.097}_{0.000}$ & $\textbf{0.323}\pm^{0.065}_{0.032}$ & 82.48\% & 11.38\%\\
8 &  $0.250\pm^{0.008}_{0.040}$ & $0.298\pm^{0.040}_{0.008}$ & $\textbf{0.306}\pm^{0.016}_{0.016}$ & 22.40\% & 2.68\%\\
\mr
Overall & $0.195\pm^{0.019}_{0.068}$ & $0.237\pm^{0.118}_{0.097}$ & $\textbf{0.331}\pm^{0.074}_{0.063}$ & 69.74\% & 39.67\%\\

\br
\end{tabular*}
\end{table}

\subsection{Evaluation Metric}
The goal of this study is to show that ML can improve beam availability and minimize damage to the accelerator by developing advanced fault prediction models and comparing their performance. 
Any anomaly prediction technique we deploy should not significantly increase the levels of beam aborts or abort duration due to false alarms. 
Hence, our evaluation metric is built to keep a fixed minimal False Positive Rate (FPR) while achieving a significant True Positive Rate (TPR). 
Even a modest level of FPR, such as 2\%, would noticeably reduce the beam availability and negatively affect the science research program.
Based on input from the operators and subject matter experts a FPR of 0.1\% is considered the upper acceptable limit.


\subsection{Model architecture selection (HPO and NAS)}
\label{ch:hpo}
Deep learning models can be built with a wide range of different configurations and architectures, producing varying results.
It is essential to explore and choose the best hyper-parameters and model architecture for a given task.
\begin{figure}[h!]
    \centering
    \begin{subcaption}{TPR at FPR of 0.1\% from various architectures of CSNN model during HPO and NAS}
        \centering
        \includegraphics[width=1.\textwidth]{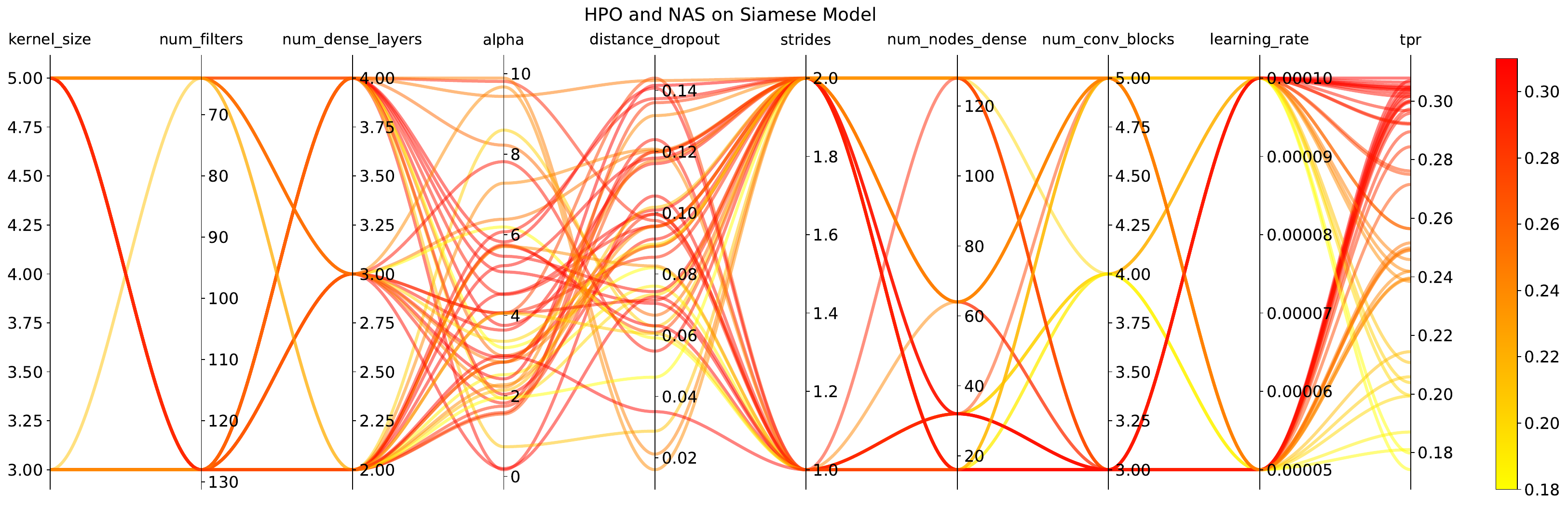}
        
    \end{subcaption}

    \begin{subcaption}{TPR at FPR of 0.1\% from various architectures of CVAE models during HPO and NAS. Note- suffix "e." and "d." represents encoder and decoder parameters respectively.}
    \centering
        \includegraphics[width=1.\textwidth]{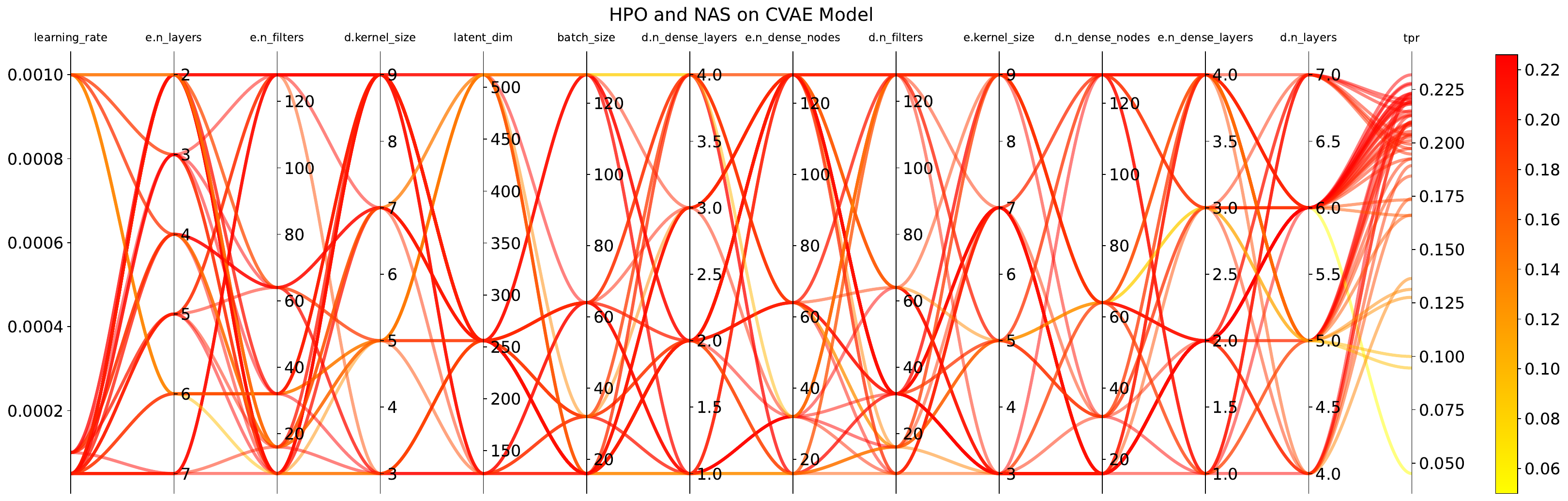}
    \end{subcaption}
    \caption{HPO and NAS study on Siamese and VAE models. TPR on validation data is used as a metric to evaluate the model performance.}
    \label{fig:HPO_NAS}
\end{figure}
There are multiple HPO and/or NAS toolkits that are available for researchers to use and produce an optimal architecture(s).
In this study, we use the Tree-based Parzen Estimator (TPE)~\cite{tpe} algorithm available in the HyperOpt~\cite{hyperopt} library in python in combination with ml-flow~\cite{mlflow} to perform HPO and NAS studies on both Siamese and VAE models in order to pick the best model architecture for the comparison.
The model initialization was fixed for this process.
The objective was set to minimize validation loss with 50 trials to be consistent with the HyperOpt usage in the literature.

In addition to the numerical hyper-parameters and model architecture parameters shown in Figure~\ref{fig:HPO_NAS}, we also explore activation functions between ReLU, and Leaky\_ReLU for both the models.
Based on this study, we compared the model architectures and hyper-parameters that produced the best TPR at the target FPR of 0.1\% in the HPO/NAS study.

\subsection{CSNN Model results and comparison to individual SNNs}

The models were trained on the above-described beam current data set from DCM with the training data generator to learn to produce a low similarity score between normal pulses collected during normal operation (named normal) and normal pulse preceding faults (named anomaly). 
\begin{figure}[h!]
    \centering
    \includegraphics[width=1\textwidth]{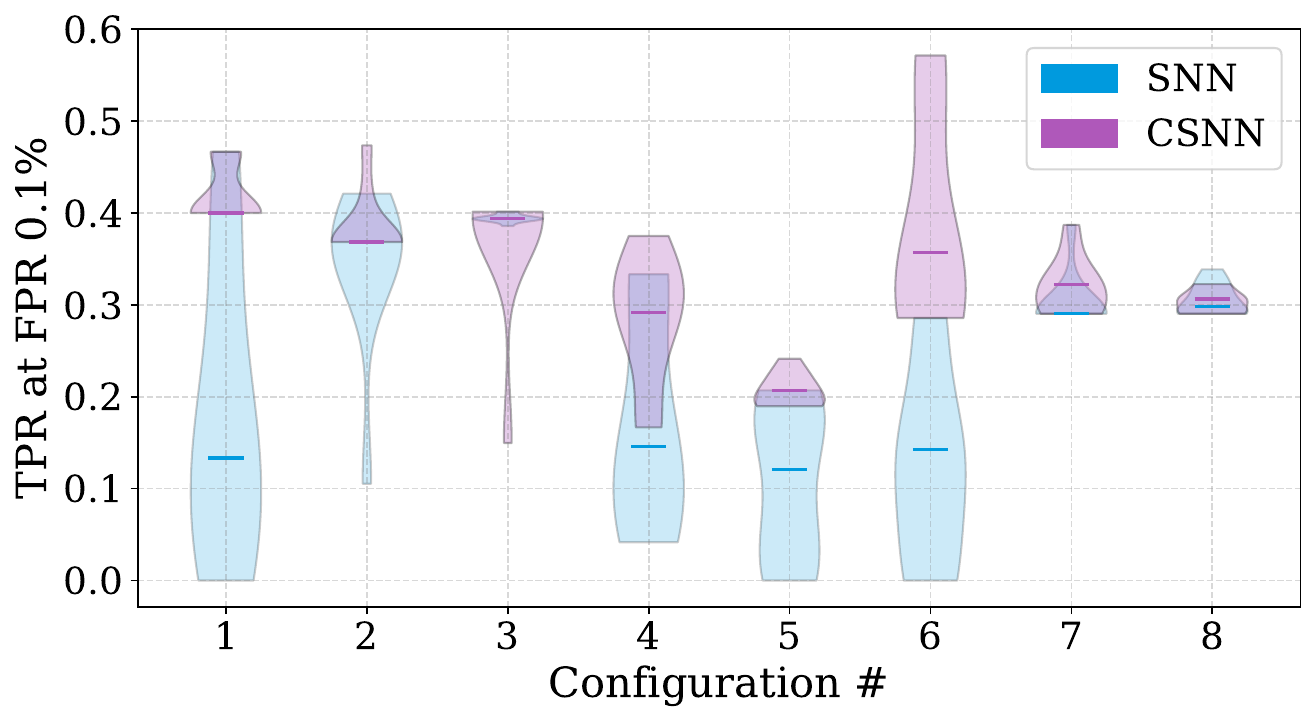}    \caption{Comparison of SNN with CSNN model. The CSNN model outperforms the single-configuration SNN model on all the beam configurations in this study. 
    }
    \label{fig:Siamese_Csiamese_Comparison}
\end{figure}
We trained eight individual SNN models on data sets belonging to each beam configuration and a single CSNN model on the entire data set using beam configurations as conditional input.
To statistically evaluate the performance, stability, and robustness of the model architecture, we trained each model (SNN and CSNN) 10 times with the same architecture (chosen after architecture optimization as described in Section~\ref{ch:hpo}), using different weight and bias initialization.
Figure~\ref{fig:Siamese_Csiamese_Comparison}, compares the prediction performance between SNN and CSNN models regarding TPR at FPR of 0.1\%.
In the plot, each blue violin represents an ensemble of SNN model inferences on the test data of the beam configuration they are trained on. 
The corresponding purple violins represent the ensemble of CSNN model inferences on the same test data.
The beam configuration ID are marked on the x-axis.
As shown in Figure~\ref{fig:Siamese_Csiamese_Comparison}, the CSNN outperforms SNN model trained on a single configuration for our application.
We believe that this improved performance is due to the CSNN model being able to leverage the correlation between beam current waveforms from different configurations and hence can learn from an increased amount of data.

\subsection{Comparision of CSNN with CVAE}
\begin{figure}
    \centering
    \includegraphics[width=1\textwidth]{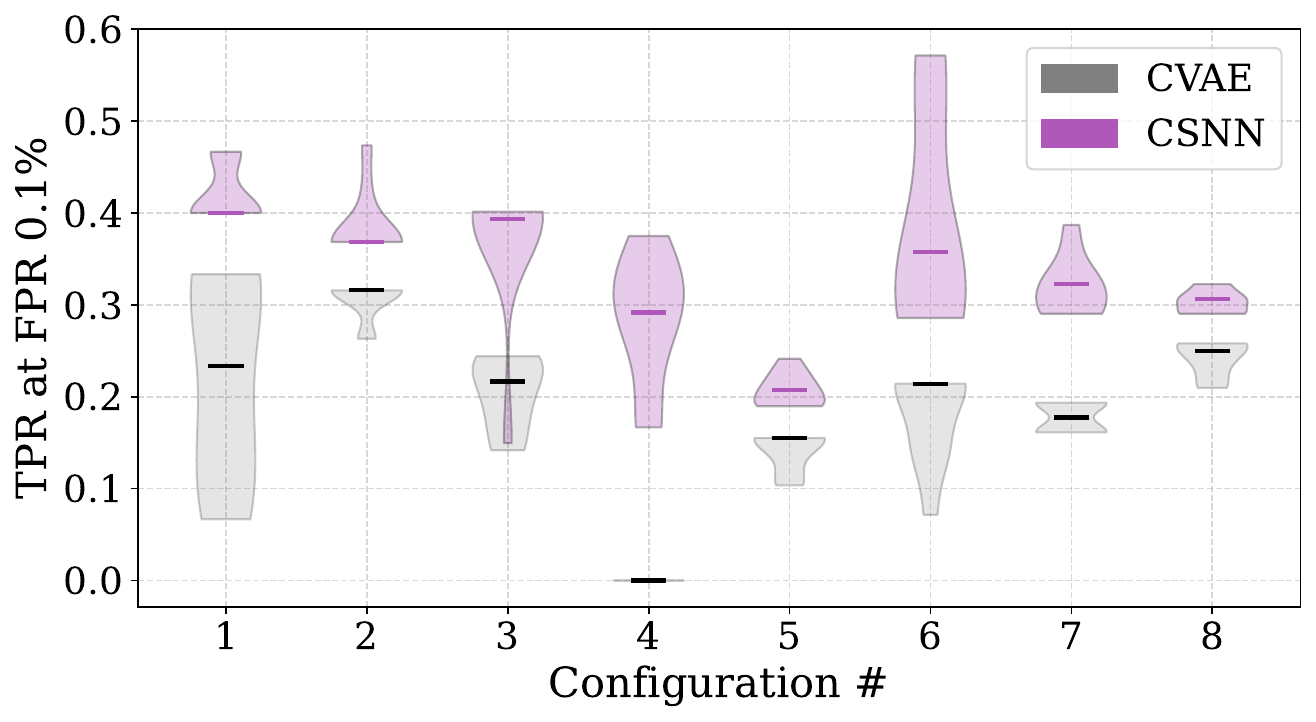}    
    \caption{Comparison of CSNN models with CVAE models. The CSNN model
outperforms CVAE on all the beam configurations}
    \label{fig:Siamese_CVAE_Comparison}
\end{figure}

To understand the stability and robustness of our models, we trained an ensemble of 10 CVAE models with different weights and bias initialization on the same data set.
Figure~\ref{fig:Siamese_CVAE_Comparison} shows the predictive results of comparing CSNN and CVAE regarding TPR at FPR of 0.1\%. 
For all the configurations, CSNN is seen to outperform CVAE.  
Even though both models, in this case, are trained on the same data set and use beam configuration as a conditional input, we believe that the CSNN model, being a supervised learning method, can leverage the label information and is much more powerful than CVAE.
In addition, the CVAE model needs to reproduce all the features to reconstruct the normal waveforms. 
In contrast, CSNN does not need to learn to reproduce the whole waveforms; instead learns to extract the relevant features to distinguish anomalies from normal.

\begin{table}
\caption{\protect\label{tab:model_architectures}CVAE and CSNN model architecture and parameters used for the results shown in this paper. Note: the numbers in the parenthesis represent the decoder configuration for the CVAE model}
\begin{tabular*}{\textwidth}{@{}l*{15}{@{\extracolsep{0pt plus
12pt}}l}}
\br
Parameter name & CSNN  & CVAE \\
\mr
Waveform input dimension & [N, 10000, 1] & [N, 10000, 1]\\
config. input dimension & [N, 8] & [N, 8] \\

Conv1D blocks & 3 & 3 (4) \\
Number of kernels & 128 & 128 (128) \\
kernel size & 3 & 6 (6) \\
Activation function & $ReLU$ & $ReLU$\\
Number of dense layers & 3 & 3 (3) \\
Unites per Dense layer & 128 & 128 (128)\\
Latent $z$ & 40 & 512\\
Optimizer & Adam & Adam\\
%
Loss & $Contrastive$ & $MSE+KLD$\\
Trainable Parameters & 250K & 2.3M \\
\br
\end{tabular*}
\end{table}

It is important to note that the two models also vary in architecture and how they learn to distinguish abnormal inputs from normal. However, deep learning models are often compared based on the number of trainable parameters they have. 
The model architectures are compared in Table~\ref{tab:model_architectures}; the number of trainable parameters in the CVAE (after a light HPO and NAS) is roughly an order of magnitude higher than that in the CSNN model. 
This indicates that the CVAE requires much more expressivity to learn to reproduce the normal data waveforms. 
On the other hand, CSNN leverages the label information and does not need to learn to reproduce the entire waveforms.
This requires much lower trainable parameters and still outperforms the CVAE model.

\section{Conclusions and Outlook}
\label{ch:conclusion}

In this paper, we have explored conditional ML models to predict errant beams with varying beam configurations at the SNS accelerator. 
We have trained and evaluated SNN individually on beam current data belonging to different beam configurations as well as CSNN and CVAE on the entire data set.
We have described the data collection, processing, methods, and training procedure along with HPO and NAS studies for optimal model architecture and parameter selection.
The comparison between the above three variations has also been presented and it can be concluded that the CSNN outperforms both SNN and CVAE by 39.67\% and 69.74\% for our application.
The CSNN model can achieve more than 30\% of TPR at our target FPR which is very promising for any application in the field. 
We can also move the threshold to adjust FPR as needed to avoid more false aborts while preventing a significant amount of errant beam pulses.
While CSNN performs better than SNN and CVAE, more studies are required to fully understand its capabilities to handle a large number of different beam configurations and adaptability in terms of continual learning.
We would also like to further explore uncertainty quantification with the CSNN model and explore 
its usage with continual learning for practical purposes. 

The conditional models presented in this paper are capable of handling sudden distribution shifts due to known configuration changes, 
however, these models have not been tested on gradual and/or sudden data drifts coming from aging machine components or other unexpected events such as field emissions. 
We would like to further develop the supervised learning approaches to better handle all the data drifts moving forward.

\ack
The authors acknowledge the help from Charles Peter and David Brown for evaluating operations requirements, Frank Liu for his assistance on the ML techniques, and Sarah Cousineau for making this grant work possible. 
This work was supported by the DOE Office of Science, United States under Grant No. DE-SC0009915 (Office of Basic Energy Sciences, Scientific User Facilities program).
This manuscript has been authored by UT-Battelle, LLC, under contract DE-AC05-00OR22725 with the US Department of Energy (DOE). 
The Jefferson Science Associates (JSA) operates the Thomas Jefferson National Accelerator Facility for the U.S. Department of Energy under Contract No. DE-AC05-06OR23177. 
This research used resources at the Spallation Neutron Source, a DOE Office of Science User Facility operated by the Oak Ridge National Laboratory. 
The US government retains and the publisher, by accepting the article for publication, acknowledges that the US government retains a nonexclusive, paid-up, irrevocable, worldwide license to publish or reproduce the published form of this manuscript, or allow others to do so, for US government purposes. DOE will provide public access to these results of federally sponsored research in accordance with the DOE Public Access Plan (http://energy.gov/downloads/doe-public-access-plan).

\section*{Appendix}
\label{Appendix_A}
In this section, we describe the effect of the conditional inputs on the input data i.e. how the beam configuration changes the beam current waveforms.
Beam current waveform can be decomposed into three sections namely, ramp-up, beam-on, and ramp-down.
The ramp-up is the beginning of the waveform when the beam current gradually increases in amplitude, beam-on is the middle section when the beam current stays high, and the ramp-down section comes at the end when the beam current gradually decreases. \\

\begin{enumerate}
    \item \textbf{Ramp Up Pulse Width} is the width of the mini-pulse at the beginning of the ramp-up section.
    \item \textbf{Ramp Up Pulse Width Change} is the period in terms of the number of mini-pulses before the width of the mini-pulse is increased by one in the ramp-up section. In other words, if it is set to 5 then the mini-pulse width is increased by one after every 5 mini-pulses.
    \item \textbf{Ramp Up Width} is the number of mini-pulses in the ramp-up section.
    \item \textbf{Beam On Pulse Width} is the width of the mini-pulse at the beginning of the beam-on section.
    \item \textbf{Beam On Width} is the number of mini-pulses in the beam-on section.
    \item \textbf{Ramp Down Pulse Width} is the width of the mini-pulse at the beginning of the ramp-down section.
    \item \textbf{Gate Source Offset} - Offset of Reference Gate relative to start of source.
    \item \textbf{Beam Reference Gate Width} - Width of the macro pulse
\end{enumerate}

%


\section*{References}
\bibliographystyle{iopart-num}
\bibliography{main, review}

\end{document}